\documentclass[pdflatex,sn-mathphys-num]{sn-jnl}

\usepackage{graphicx}%
\usepackage{multirow}%
\usepackage{amsmath,amssymb,amsfonts}%
\usepackage{amsthm}%
\usepackage{mathrsfs}%
\usepackage[title]{appendix}%
\usepackage{xcolor}%
\usepackage{textcomp}%
\usepackage{manyfoot}%
\usepackage{booktabs}%
\usepackage{algorithm}%
\usepackage{algorithmicx}%
\usepackage{algpseudocode}%
\usepackage{listings}%

\usepackage{xspace}
\usepackage{hyperref}
\let\orcidlogo\relax 
\usepackage{orcidlink}
\usepackage{graphicx}

\def\thebaroffset{0.0em}
\newcommand{\offsetoverline}[2][\thebaroffset]{\kern #1\overline{\kern -#1 #2}}%
\def\PLambda     {\ensuremath{\Lambda}\xspace}
\def\Lz          {{\ensuremath{\PLambda}}\xspace}
\def\Lbar        {{\ensuremath{\offsetoverline{\PLambda}}}\xspace}
\def\PK      {\ensuremath{\mathrm{K}}\xspace}
\def\kaon    {{\ensuremath{\PK}}\xspace}
\def\KS      {{\ensuremath{\kaon^0_{\mathrm{S}}}}\xspace}
\def\Ppi         {\ensuremath{\pi}\xspace}
\def\pion   {{\ensuremath{\Ppi}}\xspace}
\def\pip    {{\ensuremath{\pion^+}}\xspace}
\def\pim    {{\ensuremath{\pion^-}}\xspace}
\def\Pp      {\ensuremath{\mathrm{p}}\xspace}
\def\proton      {{\ensuremath{\Pp}}\xspace}

\def\to                 {\ensuremath{\rightarrow}\xspace}
\newcommand{\decay}[2]{\mbox{\ensuremath{#1\!\to #2}}\xspace}

\def\KSPIPI     {\decay{\KS}{\pip\pim}}
\def\LPPI    {\decay{\Lz}{\proton\pim}}
\newcommand{\pstar}{\ensuremath{p^{*}}}
\newcommand{\pstarsq}{\ensuremath{p^{*^2}}}
\newcommand{\estar}{\ensuremath{E^{*}}}
\newcommand{\R}{\ensuremath{R_{p}}}
\newcommand{\MOMK}{\ensuremath{p_{\KS}}}
\newcommand{\MK}{\ensuremath{m_{\KS}}}
\newcommand{\MLZ}{\ensuremath{m_{\Lz}}}
\newcommand{\MLZB}{\ensuremath{m_{\Lbar}}}
\newcommand{\MP}{\ensuremath{m_{P}}}
\newcommand{\meff}{\ensuremath{\xi}}

\begin{document}

\title[Improving systematic uncertainties on precision two-body mass measurements]{Improving systematic uncertainties on precision two-body mass measurements}

\author[1]{\fnm{Allison} \sur{Chu}\orcidlink{0009-0007-3054-6310}}\email{A.N.M.Chu@sms.ed.ac.uk}
\author[1]{\fnm{Yiming} \sur{Liu}\orcidlink{0000-0003-3257-9240}}\email{Y.Liu-293@sms.ed.ac.uk}
\author*[1]{\fnm{Matthew} \sur{Needham}\orcidlink{0000-0002-8297-6714}}\email{mneedham@ed.ac.uk}
\affil[1]{\orgdiv{School of Physics and Astronomy}, \orgname{University of Edinburgh}, \orgaddress{\city{Edinburgh}, \postcode{EH9 3FD}, \country{UK}}}

\abstract{
To make precision particle mass measurements in charged spectrometers detailed understanding of the influence of detector effects is critical. In this paper the influence of detector-related uncertainties on the determination of the parent particle mass in two-body decays is investigated. It is shown how the dependence of observed mass shifts on the sum and difference of the daughter particle momenta can be used to determine the physical causes of a bias more rigorously than the \textit{ad hoc} rules that are often adopted. The approach is illustrated using the case of measuring the $\Lambda$ hyperon mass. This observable is of interest because our current knowledge relies on information from a single experiment that has not been updated to account for changes in the value of the $\textrm{K}_{\textrm{s}}^0$ mass used for calibration. With the approach developed in the paper it shown that the LHCb experiment has the capability to make a measurement of the $\Lambda$ mass with systematic uncertainties from the tracking system controlled to $0.7\,$keV/$c^2$. This allows a total precision of $2.2\,$keV/$c^2$ to be achieved, dominated by the knowledge of the $\textrm{K}_{\textrm{s}}^0$ mass used for calibration. This would improve the current knowledge of the $\Lambda$ hyperon mass by a factor of three.
}
\keywords{Particle tracking, Detector calibration, $\Lambda$ hyperon mass}

\maketitle

\section{Introduction}
\label{sec:intro}
Accurate tracking of charged particles is a critical component of any collider physics experiment. Systematic effects, such as the momentum scale and ionization energy loss, often interplay in a non-trivial way in measurements at such experiments, especially in measurements of particle masses. Despite the importance of this topic, there is relatively little published literature in this area particularly for light resonances where the mass of daughter particles cannot be neglected. Consequently, experimentalists often rely on \textit{ad hoc} rules of thumb such as that the bias on mass measurements scales with the energy release \cite{LHCb-PAPER-2020-003,LHCb:2019epo}. A detailed understanding of these effects, their commonalities, and their potential biases, is necessary for mass measurements to be performed with the highest precision. This paper sets out pathways for correcting biases on mass measurements in a rigorous data-driven way. By formulating the corrections to the measured momentum in terms of physical parameters, such studies can improve detector modelling. Controlling systematic effects related to particle tracking is particularly important for studies at the Large Hadron Collider (LHC) due to the large size of the collected datasets. The calibration of the charged momentum scale for the LHCb experiment \cite{LHCb-DP-2008-001} in Run 1 and 2 of the LHC is described in \cite{LHCb:2020xds}. The procedure uses large samples of $J/\psi$, $\Upsilon$ and $b-$hadron decays collected during proton-proton data taking to determine the momentum scale of the detector. The relative accuracy is evaluated by considering the variation of mass with particle kinematics and across different resonances to be $3\times 10^{-4}$. This is the limiting uncertainty on many mass measurements at LHCb - for example those of open charm mesons reported in \cite{LHCb:2013fwo} and the exotic $\chi_{c1}(3872)$ state in \cite{LHCb:2020xds}. Consequently, a more accurate calibration procedure, also taking into account energy loss, is needed to improve measurements of these observables and this is one aim of this work.

The formalism developed in this paper is general and can be applied to any two-body decay. It is based upon fitting the mass in subsamples of the sum and difference of the daughter momenta. To illustrate its use, it is shown that this approach can enable a world-leading measurement of the mass of the \Lz hyperon from its decay to the $\proton\pi^-$ final state with the LHCb detector. For this mode, the topologically similar \KSPIPI decay is an ideal calibration channel, as it is more sensitive to detector biases. Experimental measurements of the \Lz mass can be compared to lattice QCD predictions~\cite{Fodor:2012gf}. In addition, comparing the masses of the $\Lz$ and $\overline{\Lz}$ hyperons provides a test of \textit{CPT}-symmetry~\cite{Sozzi:1087897}.

Using $\KSPIPI$ and $\LPPI$ decays for detector calibration is considered in the context of calibration studies for next generation $e^+e^-$ machines \cite{Altmann:2025feg, Madison:2022spc}. For those studies, the calibration method based on the Armenteros-Podolanski plot \cite{Podolanski1954Vevents} described in \cite{Rodriguez:2020qhf} is adopted. Using $\KSPIPI$ decays for calibration has the advantage of a larger dataset compared to more conventional standard candles such as $J/\psi \rightarrow \mu^+ \mu^-$ decays. However, more care is needed in the calibration procedure, as the mass of the daughters cannot be ignored. Moreover, in \cite{Rodriguez:2020qhf} only a multiplicative bias is considered. Due to the low momenta of the daughter pions and the small opening angle, measurements using $\KSPIPI$ decays  are more susceptible to both additive biases from energy loss and the vertex determination, respectively. The approach used in this study allows to quantify and correct such biases.

This paper is structured as follows. First, the current knowledge of the $\Lz$ mass is discussed, and it is highlighted that new measurements are needed since only data from one experiment carried out in the 1990s is used in the current world averages. Section~\ref{sec:simulation} describes the simulation used for these studies. Following this, in Section~\ref{sec:formalism} the general formalism used is detailed. In Section~\ref{sec:studies} it is shown how the \KS decay can be used to calibrate the measured momenta and allows the $\Lz$ mass to be measured with high precision at LHCb. Finally, Section~\ref{sec:multi} discusses how the formalism can be extended to multibody decay modes. 

\section{Current knowledge of the $\Lz$ mass}
\label{sec:pdg}
The 2024 Review of Particle Properties, the PDG~\cite{PDG2024},  reports $m_{\Lz}=1115.683 \pm 0.006 \, \textrm{MeV}/c^2$. This value is based on data collected in the 1990s at the Brookhaven AGS by the E766 collaboration~\cite{Hartouni:1994zg}. The PDG value is the weighted average of $m_{\Lz} = 1115.678 \pm 0.006 \,  \textrm{(stat)} \, \pm \,  0.006 \, \textrm{(syst)} \, \textrm{MeV}/c^2$ and $m_{\Lbar} = 1115.690 \pm 0.008 \, \textrm{(stat)} \, \pm \,  0.006 \, \textrm{(syst)} \, \textrm{MeV}/c^2$ reported in \cite{Hartouni:1994zg} under the assumption that the systematic uncertainties are not correlated. As the systematic uncertainty is entirely due to the knowledge of the \KS mass used to calibrate the momentum scale, this procedure is questionable. Furthermore, the uncertainty on the \KS mass has improved from $\MK = 497.671 \pm 0.031 \, \textrm{MeV}/c^2$ used in \cite{Hartouni:1994zg} to $\MK = 497.611 \pm 0.013 \, \textrm{MeV}/c^2$ quoted by the PDG \cite{PDG2024}. There is sufficient information in \cite{Hartouni:1994zg} to recalculate $m_{\Lz}$ to account for the change in $m_{\KS}$. The updated weighted average, with the assumption of fully correlated systematic uncertainties, is $m_{\Lz} = 1115.671 \pm 0.005 \, \textrm{MeV}/c^2$. This is a shift of around $2\sigma$ compared to the value quoted by the PDG.

In \cite{Hartouni:1994zg} radiative corrections are not considered. Although QED radiation is small for the \LPPI decay due to the limited phase space, it has a larger impact on the $\KSPIPI$ calibration channel. Using the simulation discussed in Section~\ref{sec:simulation} and the information on the fit procedure given in \cite{Hartouni:1994zg} the bias from ignoring radiative corrections on $\MLZ$ is estimated to be $\sim 0.001 \, \textrm{MeV}/c^2 $. 

The only other measurements of $\MLZ$ listed by the PDG, but not used for averages, are from bubble chamber experiments carried out in the 1960s and 1970s \cite{Bhowmik:1963zz,Schmidt:1965zz, London:1966zz,osti_4454111,Hyman:1972mp} and have large uncertainties.  Given the limited published data, a new measurement of $\MLZ$ is of interest. Naturally, such studies would allow for improved $CPT$ symmetry tests from a comparison of the $\Lambda$ and $\overline{\Lambda}$ mass. The PDG quantifies this via 
 \begin{equation}
     \textrm{R}_{CPT} = \frac{(\MLZ - \MLZB)}{\MLZ},
 \end{equation}
 and gives $\textrm{R}_{CPT} = (-0.1 \pm 1.1) \times 10^{-5}$ \cite{PDG2024}. Large numbers of $\Lz$ baryons are produced in high-energy $pp$ collisions at the LHC, allowing high statistical precision to be achieved by all the LHC experiments. The challenge is to control the systematic uncertainties. Due to increased computing power, track reconstruction and alignment techniques have improved since the 1990s. However, tracking detectors at the LHC have a much larger material budget than spectrometers using drift chamber technology such as E766. The forward geometry of the LHCb experiment \cite{LHCb-DP-2008-001}, where the vertex detector (VELO) extends to $z=80$~cm from the interaction point, leads to a high acceptance for long-lived particles such as $\Lz$ hyperons and is the focus of this paper. As the $\KSPIPI$ decay acts as a calibration channel, the ultimate precision in these studies is limited to $\sim 2\,$keV$/c^2$ by the $13\,$keV$/c^2$ uncertainty on $\MK$ given in \cite{PDG2024}.

\section{Simulation}
\label{sec:simulation}
Simulation samples of \LPPI and \KSPIPI decays are generated using the \textsc{RapidSim} fast simulation package \cite{Cowan:2016tnm} which is widely used in LHCb studies. \textsc{RapidSim} provides an interface that allows to simulate events using \textsc{EvtGen} \cite{Lange:2001uf} as well as \textsc{Photos} \cite{Golonka:2005pn} to simulate QED radiative corrections. The default \textsc{RapidSim} version does not have models for $\KS$ and $\Lz$ production. To have reasonable decay kinematics, a custom implementation of the differential cross-section versus transverse momentum is needed. For this the measurements made by the ALICE collaboration at a centre of mass energy, $\sqrt{s} = 13~\textrm{TeV}$~\cite{ALICE:2020jsh} and publicly available via \textsc{HepData} \cite{Maguire:2017ypu} are used. A sample of $10^8$ $\KSPIPI$ and $5 \times 10^7$ \LPPI decays was generated: the proportion allows for the different production cross-sections \cite{LHCb:2011ioc}. Given that $\sim 10^{13}$ \KS mesons are produced per $\textrm{fb}^{-1}$ within the acceptance of LHCb \cite{LHCb:2012vqi}, a sample of this size can be readily collected during LHC running. After requiring the daughter particles to be in the LHCb acceptance ($2< \eta < 5$) before and after the dipole magnet \footnote{The corresponds to an implicit lower momentum requirement of $p > 2 \, \textrm{GeV}/c$.}, the decay vertex to be within the VELO ($z< 80\,$cm) and applying loose kinematic requirements, $1.4 \times 10^7$ and $4\ \times 10^6$ events remain for the $\KSPIPI$ and \LPPI decays respectively. To simulate the detector resolution, the true \KS (\Lz) invariant mass is convolved with a Gaussian resolution function with $\sigma = 1.2\,\textrm{MeV}/c^2$ ($3.5 \,\textrm{MeV}/c^2$) respectively \cite{LHCb:2011ioc, LHCb:2014set}. With this resolution for the $\LPPI$ decay, a statistical precision on $\MLZ$ of $0.6~\textrm{keV}/c^2$ is found by fitting the distribution with a Crystal Ball function \cite{Skwarnicki:1986xj}.

\section{Formalism}
\label{sec:formalism}
\begin{figure}
    \centering
    \includegraphics[width=0.8\linewidth]{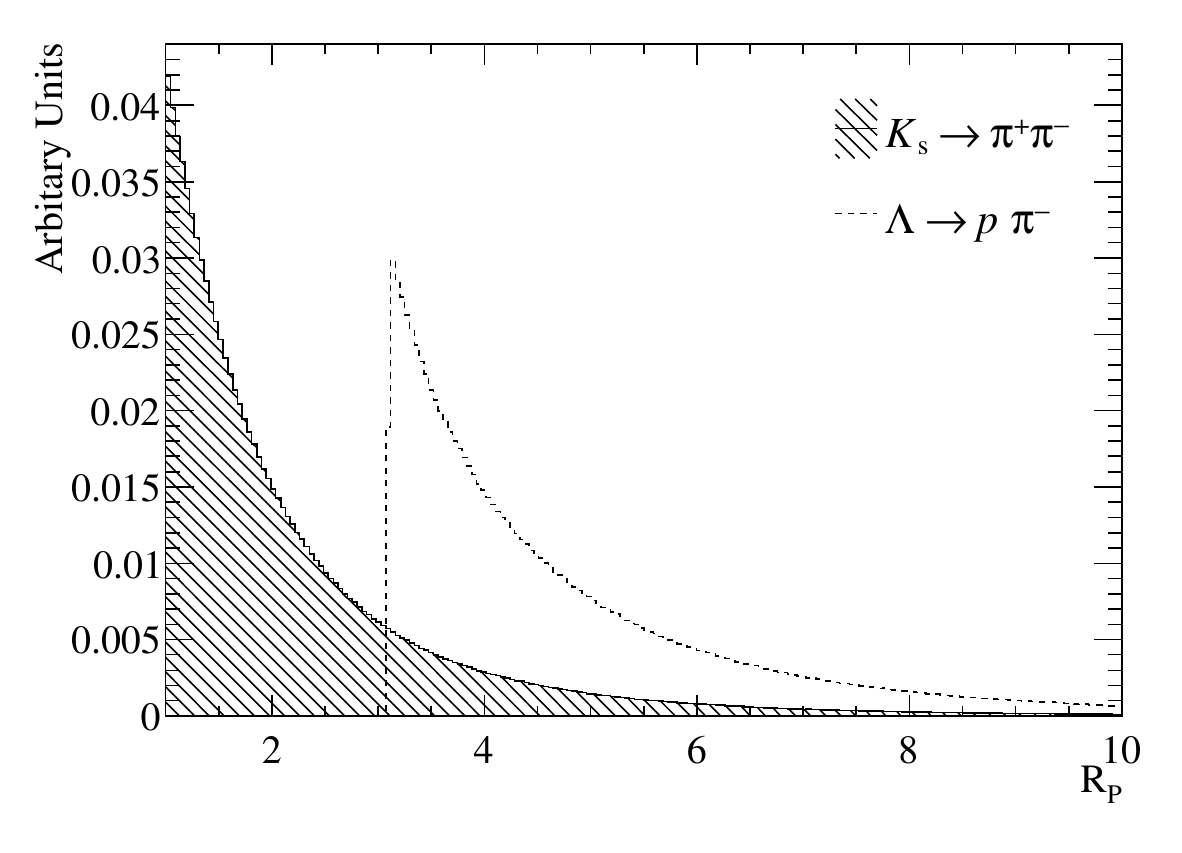}
    \caption{Distribution of the \R\ variable, taken to be the maximum of $p_1/p_2$ and $p_2/p_1$  for the two \KSPIPI and \LPPI simulation samples. }
    \label{fig1}
\end{figure}

The formalism presented here extends that described in \cite{LHCb:2023ood}. For a two-body decay $P \rightarrow d_1d_2$,  the invariant mass
of the system, in natural units, is  
\begin{equation}
  \MP^2 =  m_1^2 + m_2^2  + 2 (E_1 E_2 - \vec{p}_1 \cdot \vec{p}_2)  ,
  \label{eq:one}
\end{equation}
where $m_{1,2}$, $\vec{p}_{1,2}$, $E_{1,2}$ are the mass, the momentum vector, and the energy of the decay products. In the relativistic limit, $m \ll p$ for the daughters of $\KS$ and $\Lz$ decays, and hence $E_i \approx p_i \cdot (1+m_i^2/(2 \cdot p_i^2))$. Consequently, Equation~\ref{eq:one} becomes 
\begin{equation}
 \MP^2 \approx (1+p_2/p_1) m_1^2 + (1+ p_1/p_2) m_2^2 + 2 p_1p_2 (1- \textrm{cos}\theta),
  \label{eq:two}
\end{equation}
where $\theta$ is the opening angle between the daughter particles. This form highlights the dependence on the daughter-particle momentum asymmetry, $\R = p_1/p_2$. As can be seen in Figure~\ref{fig1}, since $m_{\proton} > m_{\pim}$, for \LPPI decays the proton momentum in the laboratory frame is larger than the pion while for 
 \KSPIPI decays the two pions have more equal momenta and hence tend to have $\R \approx 1$. For the \LPPI decay, the minimum value of $\R$ is given by $R^{\textrm{min}}=(\beta \estar_{\proton} - \pstar)/(\beta \estar_{\pi^-} + \pstar)$, where $\pstar$ is the momentum of the daughter particle in the \Lz rest-frame and $\beta = v/c$ is the Lorentz boost to the laboratory frame. In the relativistic limit $\R^{\textrm{min}} \approx 3$ for the  \LPPI decay mode. 

Biases on the reconstructed mass arise in several ways and are evaluated by considering the derivatives of Equation~\ref{eq:two} with respect to $p_i$ and $\theta$. One possibility, as detailed in \cite{LHCb:2023ood} is a bias on the momentum scale, $p_i \rightarrow (1 + \alpha) p_i$. This could arise if the integrated field is incorrectly measured. If $\alpha$ is the same for both daughters, the observed bias on the parent mass is
\begin{equation}
  \Delta \MP \approx \frac{\alpha}{m_P} \cdot \: \left( K - m_1^2 \frac{p_2}{p_1} - m_2^2 \frac{p_1}{p_2} \right),
  \label{eq:alpha}
\end{equation}
where $K=m_P^2 - m_1^2 - m_2^2$. This can be written as $\Delta {m_P} = \alpha \cdot \meff$, where $\meff$ is an effective mass that depends on the masses of the daughter particles and $\R$. If the $\KSPIPI$ decay is used to determine $\alpha$ then the correction to the $\Lz$ mass is
\begin{equation}
\Delta m_{\Lz} \approx \frac{\meff_{\Lz}}{\meff_{\KS}} \cdot \Delta m_{\KS}. 
\end{equation}
Using the mean values of $\meff_{\KS}$  and $\meff_{\Lz}$  obtained from the simulation samples described in Section~\ref{sec:simulation} this evaluates to $\Delta m_{\Lz} \sim 0.15 \cdot \Delta m_{\KS}$. This illustrates the power of a $\KS$ calibration sample to control momentum scale uncertainties on the measured $\Lz$ mass. An important subsample for calibration studies is $\KS$ decays with $0.91 < \R < 1.1$ (Section~\ref{sec:studies}) for which $\Delta m_{\KS} = \alpha \cdot (m_K^2- 4 m^2_{\pi^-})/m_{\KS}$. Noting
\begin{equation}
p^* = \frac{\sqrt{m_{\KS}^2- 4 m^2_{\pi^-}}}{2},
\end{equation}
Equation~\ref{eq:alpha} simplifies to
\begin{equation}
  \Delta \MK \approx \frac{4 \cdot \alpha \cdot \pstarsq}{\MK}.
\label{eq:alpha2}  
\end{equation}
Though this formula has been derived for the case of the $\KSPIPI$ decay, it can be applied to any two-body decay with equal mass daughter particles. Since the PDG lists both parent particle masses and $p^*$, for any given value of $\alpha$ the expected mass shift is readily calculated. A rule of thumb that is sometimes used experimentally,  for example in \cite{LHCb-PAPER-2020-003,LHCb:2019epo}, is that $\Delta \MP = \alpha Q$ where $Q = \MP - m_1 -m_2$ is the energy release. From Equation~\ref{eq:alpha} it is clear that though the bias on the mass is larger for decays with higher energy release, the assumption of direct scaling with the $Q$-value is in general incorrect. This is easily seen for symmetric $\KSPIPI$ decays using Equation~\ref{eq:alpha}. If $\alpha$ is known, scaling using the $Q$-value  underestimates the mass shift by a factor $1+2m_{\pi^{\pm}}/\MK \sim 1.6$.

A second possibility is a bias from the correction for energy loss in the detector material made in the track fit. In this case, $p_i \rightarrow p_i + \delta_i$. The value of $\delta$ is given by the Bethe equation \cite{PDG2024}. It depends on the material traversed and the $\beta \gamma$ of the particle. In the relativistic limit, $\delta$ increases slowly with $p_i$ and is reasonably well approximated by 
\begin{equation}
    \delta_i = a_0 \cdot t \cdot (a_1 + a_2 \cdot \textrm{log}(p_i)),
    \label{eq:paramB}
\end{equation}
where $t$ is the thickness of the material and the parameters $a_i$ depend on both the material and particle type \footnote{Parameterizing in this way is convenient as $a_0$ only depends on the properties of the material and hence can be more easily related to the radiation length.}. The logarithmic dependence on the momentum means that $\delta$ is often taken to be constant, especially at high momentum. The bias on the parent mass from energy loss is
\begin{equation}
\Delta \MP \approx \frac{1}{2 \: \MP} \cdot \left[ \delta_1\left( \frac{ K- 2 m_1^2 p_2 / p_1}{ p_1 } \right) + \delta_2 \left( \frac{ K- 2 m_2^2 p_1 / p_2}{ p_2 } \right) \right],
\label{eq:delta}
\end{equation}
implying the mass bias due to energy loss decreases with increasing daughter momentum. For a forward detector such as LHCb, energy loss has a larger impact on $\KS$ decays compared to heavier resonances such as the $\Upsilon$ resonances or $Z^0$.  As in the case of the momentum scale bias, the $\KSPIPI$ decay mode is more sensitive to energy-loss than the $\LPPI$ decay, but the dependence is different. Using the simulation sample described in Section~\ref{sec:simulation} it is found that  $\Delta m_{\Lz} \sim 0.3 \cdot \Delta m_{\KS}$  (\textit{cf.} 0.15 for the ratio in the case of a momentum scale bias). This illustrates an important point. Imagine a bias on the $\KS$ mass is observed in data. Fitting the full $\KSPIPI$ sample, it is possible to determine and correct for a momentum scale bias. However, if the bias is in fact due to energy loss, the determined correction and assigned systematic uncertainty will be too small. That is, in applying a calibration derived from a control mode to other modes, a good understanding of the physical cause of the bias is needed.\footnote{This also highlights that the use of several control modes with different kinematics is a powerful way to control systematic effects.} Equation~\ref{eq:delta} simplifies for $\KS$ decays with $\R \approx 1$ where $p_1 = p_2 = \MOMK/2$ and $\delta = \delta_1 = \delta_2$ to 
\begin{equation}
\Delta m_{\KS} \approx 2\cdot \frac{\delta}{\MOMK} \cdot \frac{\MK^2- 4 m^2_{\pi^-}}{\MK} = \frac{8\pstarsq\delta}{\MK \cdot \MOMK}.
\end{equation}

Another possibility is that the opening angle between the two particles is biased such that  $\theta \rightarrow \theta + \Delta \theta$. This can arise if the daughter particles are not sufficiently separated at the first measurement point to give distinct clusters in the vertex detector. It is straightforward to evaluate the derivative of Equation~\ref{eq:one}  with respect to $\theta$ and determine
\begin{equation}
\Delta \MP = \frac{p_1 p_2 \cdot \textrm{sin} \theta \cdot \Delta \theta}{\MP}.
\end{equation}
For a $\KS$ decay with $\R \approx 1$ and small $\theta$, equation~\ref{eq:two} gives
\begin{equation}
\theta \approx \frac{2\sqrt{\MK^2 - 4 m_{\pi}^2}}{\MOMK},
\end{equation}
and hence
\begin{equation}
\Delta \MK \approx  \frac{\sqrt{\MK^2 - 4 m_{\pi}^2} \cdot \MOMK \cdot \Delta \theta}{2\MK} = \frac{\pstar \cdot \MOMK \cdot \Delta\theta}{\MK}.
\end{equation}
Since the distribution of $\textrm{cos}\theta$ is similar for the $\LPPI$ and $\KSPIPI$ decays, the mass bias is comparable in magnitude for the two modes. Hence, it is important to correctly determine $\Delta \theta$ in the calibration procedure.

All the biases discussed above may be present in the data. For a $\KS$ decay with $\R \approx 1$ the total bias is given by
\begin{equation}
\Delta \MK  \approx \frac{4\pstarsq}{\MK} \cdot  \left( \alpha +\frac{2\delta}{\MOMK} +\frac{\MOMK}{4\pstar}\Delta\theta \right)
\label{eq:master}
\end{equation}
in the relativistic limit. This equation is applicable to any symmetric two-body decay to daughter particles with equal mass. It has not to our knowledge been written down in the literature previously. The possibility of biases from energy loss or mismeasurement of the opening angle were not considered in the  Armenteros-Podolanski plot method described in \cite{Rodriguez:2020qhf}. It is beyond the scope of this study, but such biases likely complicate that approach as they distort the half-ellipse on which the method relies rather than giving a simple scaling. 

In addition to the biases above for asymmetric decays. where $\R$ is not close to unity, curvature biases due to 
to detector misalignment can be important~\cite{Amoraal:2012qn}. For a forward spectrometer, such as LHCb, a misalignment that leads to a curvature bias is a small displacement, in the bending plane, of the detectors upstream and downstream of the magnet. Defining the curvature of the particle $i$ as $\omega_i = q_i/p_i$, where $q_i$ is the particle charge, the effect of a curvature bias $\omega_i \to \omega_i + \Delta\omega$ on the two-body invariant mass is
\begin{equation}
\Delta \MP \approx \frac{-\Delta \omega}{2 \: \MP} \cdot \left[ q_1 \cdot p_1\left( K- 2 m_1^2 \frac{p_2}{p_1}  \right) + q_2 \cdot p_2 \left( K- 2 m_2^2 \frac{p_1}{p_2} \right) \right].
\label{eq:curv}
\end{equation}
For \KSPIPI decays, Equation~\ref{eq:curv} simplifies to $\Delta \MP= -\Delta \omega \cdot \MK \cdot (p_1-p_2)/2$. Since the derivatives with respect to $p_1$ and $p_2$ have opposite signs, decays with $\R \approx 1$ are unaffected by a curvature bias. In contrast, the reconstructed mass for asymmetric decays is biased, and this bias increases with the magnitude of the difference in the daughter momenta \cite{LHCb:2023ood}. The change in the sign of the bias for events with $p_1 > p_2$ compared to $p_2<p_1$ has an interesting consequence. For the \KSPIPI decay, where $m_1 = m_2$, if the detector acceptance is the same for particles with positive and negative charge, the average mass of the full sample remains unbiased in the presence of a curvature bias. However, the asymmetry of the decay $\LPPI$ means that this cancellation occurs only when the decays of $\Lz$ and $\overline{\Lz}$ are considered. Therefore, in \textit{CPT} tests, care is needed to ensure the absence of a curvature bias. One way to reduce the impact of a curvature bias is to reverse the polarity of the magnetic field, as from Equation~\ref{eq:curv}  this gives an additional cancellation. Although a curvature bias will not bias the mean mass value for symmetric decays, it will degrade the resolution, particularly for high-mass states such as the $\Upsilon$ resonances and the $Z^0$ boson \cite{LHCb:2023ood} where the daughter particles have momenta in the range of 0.1-1 TeV. The sensitivity of the $Z^0$ sample to this effect is used in \cite{LHCb:2023ood} to determine the curvature bias using a pseudo-mass method. 

The proposed calibration strategy is then as follows. A fit of Equation~\ref{eq:master} to $\Delta \MK$ in subsamples of $\MOMK$ for symmetric $\KS$ decays determines the parameters $\alpha, \delta$ and $\Delta \theta$. In a second step, the presence of a curvature bias is quantified by fitting the mass in subsamples of the momentum difference for the full sample of $\KSPIPI$ decays. 

\section{Quantifying and correcting biases: determining the $\Lz$ mass}
\label{sec:studies}
\begin{figure}
    \centering
    \includegraphics[width=0.8\linewidth]{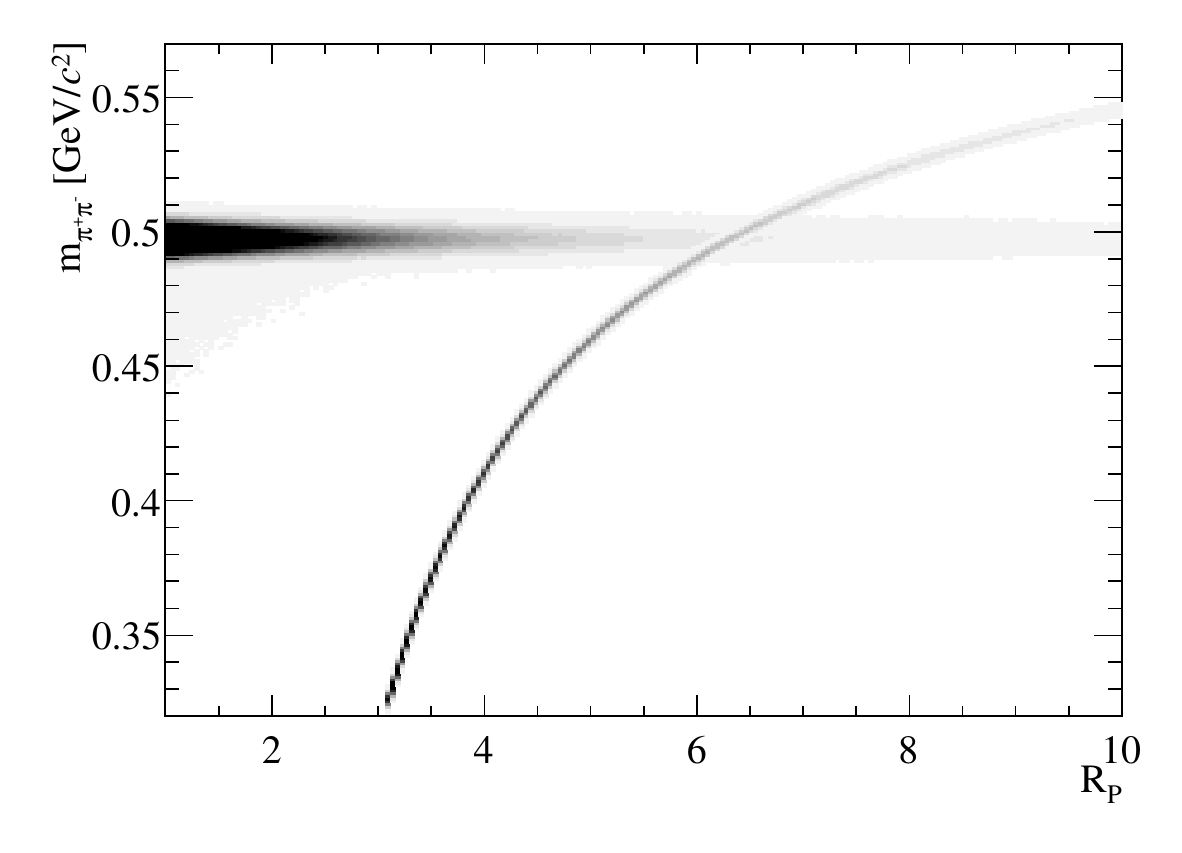}
    \caption{Invariant mass calculated under the two-pion mass hypothesis for the two simulation samples versus \R, taken here to be the maximum of $p_1/p_2$ and $p_2/p_1$. Using the $\pi^+\pi^-$ hypothesis, $\LPPI$ decays give a characteristic curved band versus $\R$. This illustrates how the two decays can be separated kinematically using $\R\,$ even if particle identification information is unavailable. }
    \label{fig2}
\end{figure}

The accuracy of the calibration procedure proposed in Section~\ref{sec:formalism} is studied using the simulation described in Section~\ref{sec:simulation} for the example of the LHCb detector. Two illustrative studies are performed to validate the method using publicly available information on detector performance. For these studies, reasonable ranges for the parameters $\alpha$, $\delta$, $\Delta \theta$ and $\Delta \omega$ are chosen based on available information on the performance of the current LHCb alignment and calibration procedures (Table~\ref{tab:parameters}). Reference \cite{LHCb:2014set} states that the LHCb detector material is known with a relative precision of $10 \, \%$. From \cite{Fave:2008zz} a particle sees on average a $35\,\%$ of a radiation length ($X_0$) of material before the LHCb dipole magnet. Multiplying these values gives the value $\Delta X_0 = t/X_{0}^{\textrm{Si}}  = 3.5 \, \%$ listed in Table~\ref{tab:parameters}. To convert from $X_{0}^{Si}$ to $\delta$ the detector material is taken to be silicon \footnote{Aluminium gives similar results.} and the expected energy loss is generated using the Bethe formula and the information on material properties in ~\cite{PDG2024,Sternheimer:1983mb}. In the calibration procedure, Equation~\ref{eq:paramB} is used. To fit the $\KS$ and $\Lz$ invariant-mass distributions, a Crystal Ball function is used \cite{Skwarnicki:1986xj}. In this way, the impact of radiative corrections is taken into account following the procedure described in \cite{LHCb:2023ood}. In these illustrative studies, the influence of background is ignored. The relatively long lifetime of the $\KS$ means that the combinatorial background is small. In a more realistic study, the background of $\LPPI$ decays needs to  be considered. Though background from $\LPPI$ decays does not have an impact when selecting candidates with $\R \approx 1$ (Figure~\ref{fig2}) it must be vetoed or included as a component in fits to the full sample.  
\begin{table}[htb!]
   \centering
 \caption{ Ranges considered on the calibration parameters. In pseudoexperiments each parameter is sampled from a Gaussian distribution with mean zero and the $\sigma$ specified below. In the absence of available information, a reasonable range for $\Delta \theta$ is chosen. \label{tab:parameters}}
 \begin{tabular}{lcc} \hline
   \toprule
       Parameter  & $\sigma$  & Source \\
       \midrule
       $\alpha$   & $3 \times 10^{-4}$ & Accuracy reported in \cite{LHCb:2023ood}\\  
        $\Delta X_0$   & $3.5 \, \%$ & Information from \cite{LHCb:2014set,Fave:2008zz} \\
        $\Delta \theta$ & $5 \times 10^{-6} \,\textrm{rad}$ & --- \\
        $\Delta \omega$ &  $ 5 \times 10^{-5} \, \textrm{c/GeV}$ & Accuracy reported in \cite{LHCb:2023yqm} \\
        \botrule
    \end{tabular}
\end{table}

In the first study, it is assumed that curvature biases are negligible ($\Delta \omega = 0$) because they are determined from studies of the $Z^0$ boson or reduced by reversing the polarity of the dipole magnet. One hundred pseudoexperiments are generated, each with a random set of values of $\alpha, \delta$, and $\Delta \theta$ generated using the values in Table~\ref{tab:parameters}. 

\begin{figure}
    \centering
    \includegraphics[width=\linewidth]{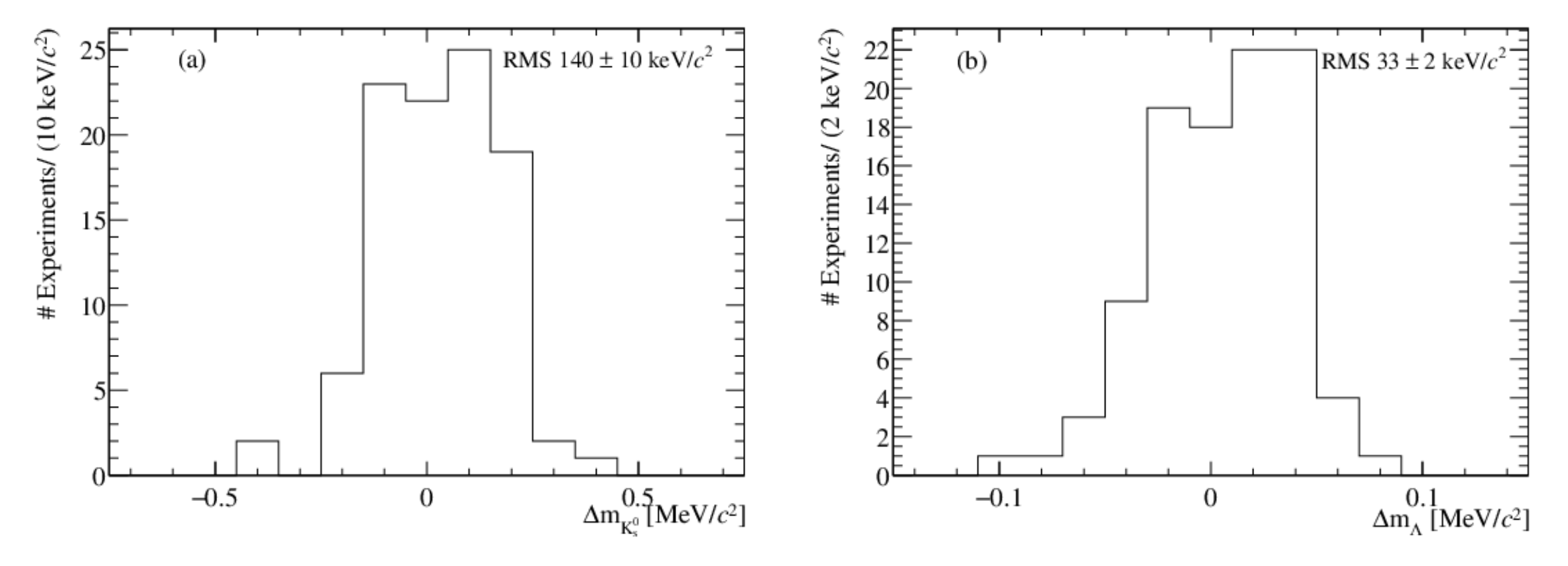}
    \caption{(a) Distribution of $\Delta \MK$ in pseudoexperiments determined from a fit of a Crystal Ball function to the full sample. (b) Distribution of $\Delta \MLZ$ in pseudoexperiments from a Crystal Ball function fit. Note the change in scale on the $x-$axis.  }
    \label{fig3}
\end{figure}

For calibration, symmetric $\KSPIPI$ decays with $0.91<\R<1.1$ are selected. This corresponds to $\sim 10 \, \%$ of the events in the $\KS$ sample and is large enough to determine with high precision without introducing large biases from the assumptions made in Equation~\ref{eq:master}. The window on $\R$ is chosen to be slightly asymmetric so that the average value of the sample is one. Fitting the $\KS$ and $\Lz$ invariant mass distributions for these pseudoexperiments gives the distributions in Figure~\ref{fig3} for the mass bias. The RMS of these distributions reflects the scale of the systematic uncertainty associated with the standard LHCb calibration procedure. As expected, the RMS of the distribution for the $\Lz$ mass is a factor of 4-5 less than for the $\KS$ case. 

\begin{figure}
    \centering
    \includegraphics[width=\linewidth]{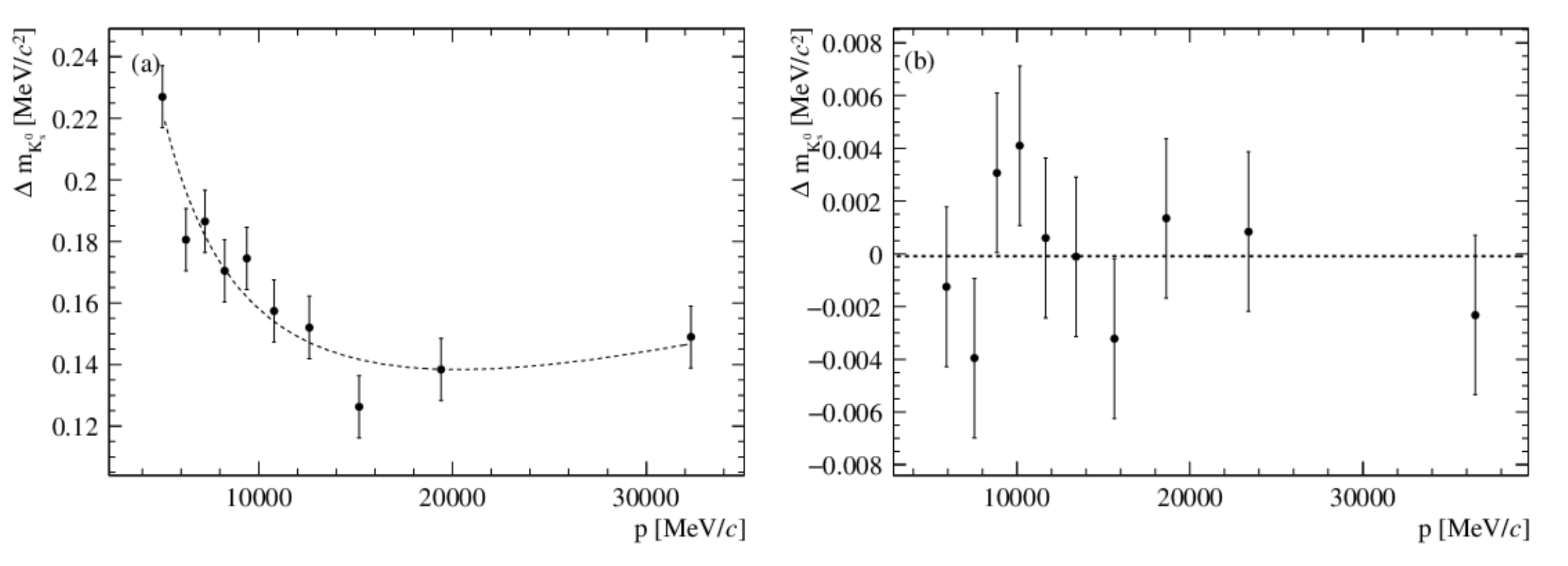}
    \caption{Distribution of $\Delta \MK$ in an example pseudoexperiment with $\alpha = 1.5 \times 10^{-4}, \delta = 2.6 \, \textrm{MeV}$ and $\Delta \theta = 5 \times 10^{-6} \, \textrm{rad}$. (a) Before calibration, a fit to Equation~\ref{eq:master} is superimposed. (b) After calibration, a fit to a constant (which has a probability of $\chi^2$ of the fit is 0.66) is superimposed. Note the change of scale on the $y-$axis. }
    \label{fig4}
\end{figure}

To determine the calibration parameters $\alpha, X_0$ and $\Delta \theta$, the symmetric $\KS$ data set is divided into ten subsamples of $\PK$ each with an equal number of entries. A fit of a Crystal Ball function is made to each subsample. Fitting Equation~\ref{eq:master} to the resulting $\Delta \MK$ values determines the calibration parameters. In a second step, the overall scale is fixed to $\MK$ by fitting the full sample with no requirement on $\R$. Figure~\ref{fig4} shows $\Delta \MK$ versus $\MOMK$ for an example pseudoexperiment before and after the calibration procedure. The model given by Equation~\ref{eq:master} describes the simulated data well. Table~\ref{tab:bias} summarizes the bias and spread of difference between the input and output calibration parameters. Although there are biases on the fitted parameters themselves, the bias on the $\Lz$ mass calculated using the parameters is small. A bias of $0.1\,\textrm{keV}/c^2$ with a standard deviation of $0.2\,\textrm{keV}/c^2$ is found in the pseudoexperiments. These should be compared to the statistical uncertainty on $\MLZ$ of $0.6 \, \textrm{keV}/c^2$ and the $2 \, \textrm{keV}/c^2$ uncertainty from the knowledge of $\MK$. 

\begin{table}[htb!]
    \centering
    \caption{Average and standard deviation of the difference between the input and output calibration parameters determined from the 100 pseudoexperiments. \label{tab:bias}}
    \begin{tabular}{lcc}
      \toprule
      Parameter  &  Average  & Standard Deviation \\
      \midrule
$\alpha$         & $-(1.32 \pm 0.03) \times 10^{-5}$ & $(3.0 \pm 0.2) \times 10^{-7}$\\ 
$X_0 $         & $(-0.46 \pm 0.07)\times 10^{-3}$ &  $(0.7 \pm 0.1) \times 10^{-3}$\\
$\Delta \theta$ [rad]        & $(4.6 \pm 0.1)\times 10^{-7}$ & $(1.3 \pm 0.1) \times 10^{-7}$  \\
\botrule
 \end{tabular} 
\end{table}

In the second, more extended study, the impact of a curvature bias is also studied. The first step of the calibration procedure to determine $\alpha, X_0 $ and $\Delta \theta$ in the same way as the study described above. To determine $\Delta \omega$, the mass is recalculated with the values obtained for these parameters and the fits are made in subsamples of the momentum difference. A linear fit of the resulting values of $\Delta \MK$ determines $\Delta \omega$ (from the slope) and the global scale of the full sample (from the offset). This procedure is iterated once. 

\begin{figure}
    \centering
    \includegraphics[width=\linewidth]{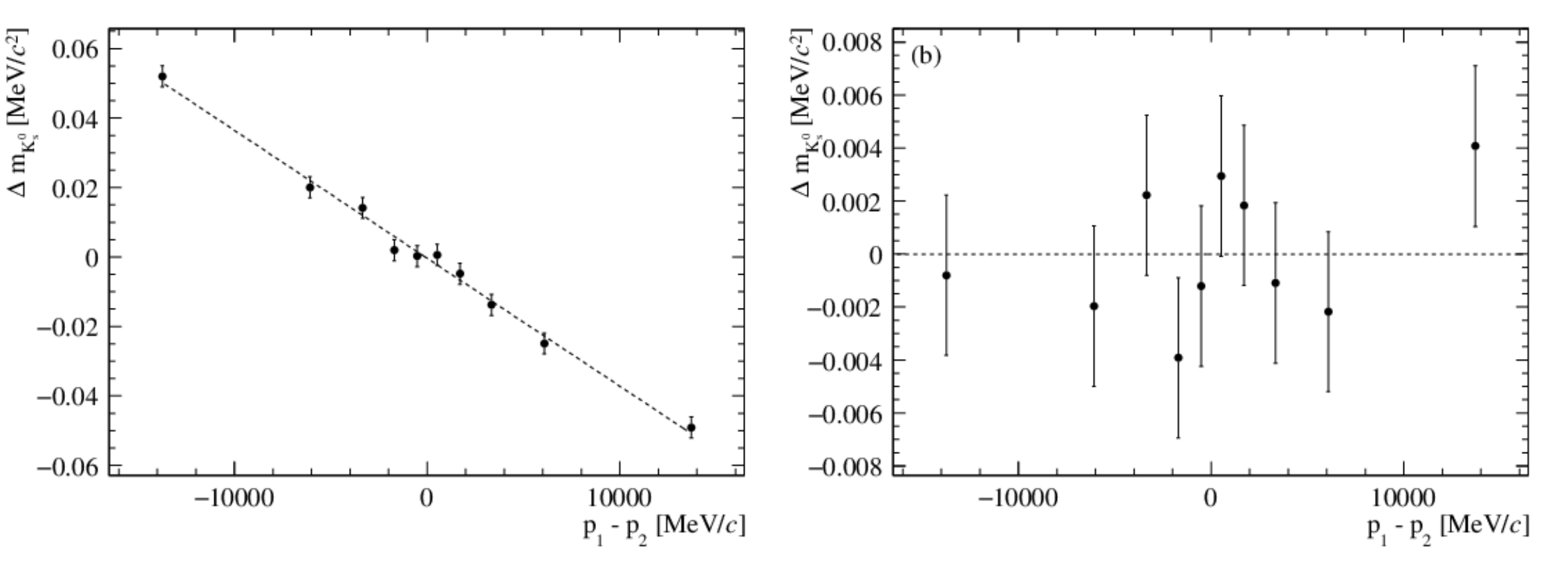}
    \caption{Distribution of $\Delta \MK$ in an example pseudoexperiment with $\alpha = 1.5 \times 10^{-4}, \delta = 2.6 \, \textrm{MeV}$, $\Delta \theta = 5\times 10^{-6} \, \textrm{rad}$ and $\Delta \omega = -1.5 \times 10^{-5} \textrm{c/GeV}$. (a) After the first calibration step. A fit to linear form is superimposed (b) After the full calibration procedure. A fit to a constant is superimposed. The probability of $\chi^2$ for the fit is 0.66. Note the change of scale on the $y-$axis. }
    \label{fig5}
  \end{figure}

Figure~\ref{fig5} shows $\Delta \MK$ versus the momentum difference of the daughter particles after the first step and the full calibration procedure. Post-calibration, the bias on the mass is consistent with being flat versus the mass difference. The bias obtained across the 100 pseudoexperiments is similar to the first study ($0.2\,\textrm{keV}/c^2$) but the uncertainty determined from the standard deviation across the pseudoexperiments increases to $0.7\,\textrm{keV}/c^2$. Before calibration, the standard deviation of $\MLZ - \MLZB$ is $30 \, \textrm{keV}/c^2$, which highlights the importance of correcting curvature biases for the $CPT$ tests. After calibration, the standard deviation reduces to $1.3\, \textrm{keV}/c^2$ and there is a bias of $0.6\,\textrm{keV}/c^2$. This study indicates that it is possible to probe $R_{CPT}$ to the $10^{-6}$ level at LHCb. This would be an order of magnitude improvement compared to current knowledge. Better control of the systematic uncertainty from the knowledge of $\Delta \omega$ can be achieved by using the $Z^0$ sample as in \cite{LHCb:2023yqm}, reversal of the magnetic field and selecting regions of phase space where the detector acceptance is equal for particles of opposite charge. 

\section{Extensions of the method}
\label{sec:multi}
In addition to the effects discussed in this paper, other biases need to be considered in a realistic study. Decays-in-flight and hadronic interactions in the detector give asymmetric tails in the resolution function. As described in \cite{LHCb:2025gog}, these can be rejected using muon detector and track quality information. A particular challenge for measurements at the LHC is the high-multiplicity environment, which means that hits from other particles might be wrongly assigned to the reconstructed trajectory close to the decay-vertex leading to a biased opening angle measurement.
Physics at the LHC is focused on short-lived particles with decay vertices close to the interaction point. Consequently, the track reconstruction may make assumptions that lead to biases for longer-lived particles necessitating a calibration as a function of the decay-vertex position. Biases of this type are discussed in the context of the ALICE experiment in \cite{alicemsc} and would need to be quantified in a realistic study. 
Finally, during the LHCb track reconstruction, particle identification information is not available. Hence, the applied energy-loss correction assumes that all particles are pions \cite{LHCb:2023ood}. The impact of this assumption for the case of the LHCb detector is evaluated using the simulation described above. With $\sim 35\,\% \, X_0$ of silicon before the dipole magnet, the bias on the reconstructed proton momentum is $\sim2$\,MeV$/c$, largely independent of the proton momentum. This momentum bias causes the reconstructed $\Lz$ mass to be shifted by $6 \, \mathrm{keV}/c^2$. Given the size of the bias compared to the achievable statistical uncertainty, it needs to be corrected for.

Linking observed mass shifts to calibration parameters is more complicated for multibody decays, particularly for biases related to the measurement of opening angles. The procedure adopted will depend on the decay mode and its resonant structure. As an illustrative example, the decay $\psi(2S) \rightarrow J/\psi \pi^+ \pi^-$ with the subsequent decay $J/\psi \rightarrow \mu^+ \mu^-$  is considered. This mode has a clear experimental signature and well undertstood dipion mass spectrum. Large samples of this decay have been collected at the LHC. Since the decay chain contains two low momentum pions, this channel is particularly useful to calibrate the energy loss correction. In addition this decay has a similar topology to the $\chi_{c1}(3872) \rightarrow  J/\psi \pi^+ \pi^-$ decay and acts as an important control channel for studies of the $\chi_{c1}(3872)$ line shape \cite{LHCb:2020xds}.

\begin{figure}
  \centering
 
    \includegraphics[width=0.98\linewidth]{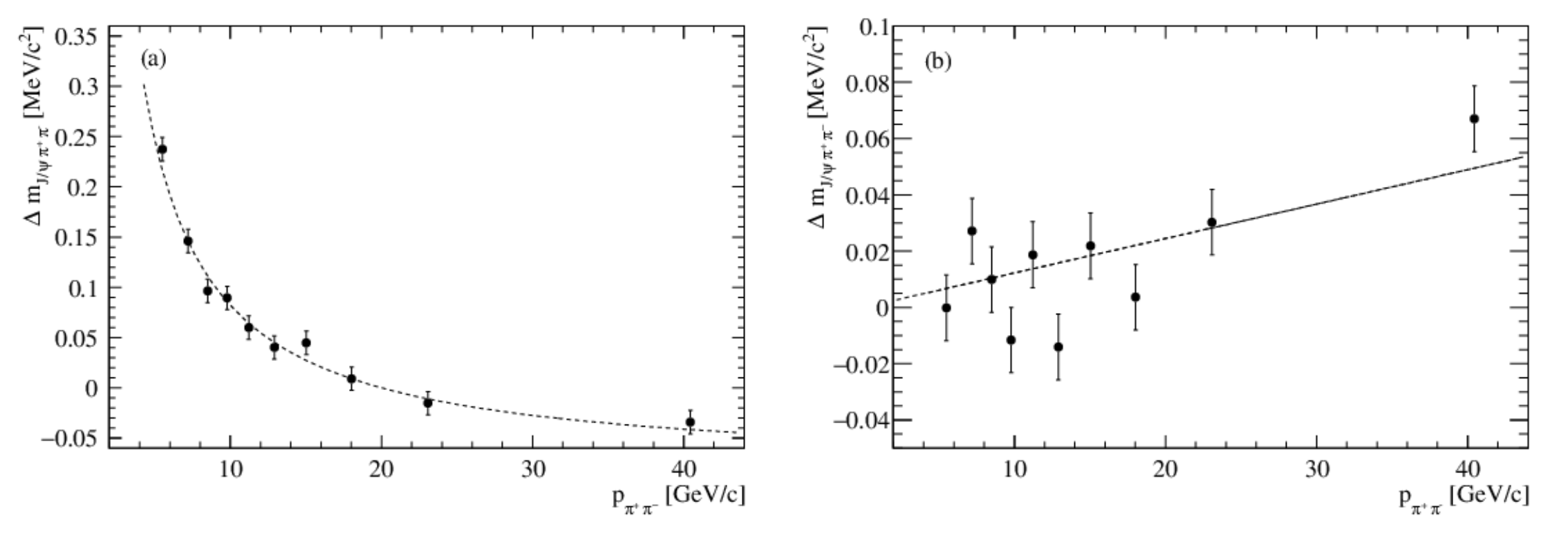} 
    \caption{(a) Mass shift, $\Delta m_{\psi(2S)}$ verus $p_{\pi^+\pi^-}$. The points are a simulation sample generated with \textsc{RapidSim}. In the simulation, the values $\alpha = 2 \times 10^{-4}$ and $\delta = 2 \, \textrm{MeV}/c$ were used. The dotted line shows the expectation for the mass shift with these values and the functional form given in Equation~\ref{eq:psi}. (b) Mass shift,  $\Delta m_{\psi(2S)}$ versus $p_{\pi^+\pi^-}$ for a simulation sample with the opening angle between the pion pair increased by $5 \times 10^{-6}\mu$rad. The resulting bias is compared to the linear form expected from Equation~\ref{eq:psi2}.  }
    \label{fig6}
  \end{figure}

In studies of $\psi(2S) \rightarrow J/\psi \pi^+ \pi^-$  at LHCb a kinematic fit of the decay tree \cite{Hulsbergen:2005pu} is used to constrain the mass of the dimuon pair to the known $J/\psi$ mass. Consequently, to good approximation, only the pions are affected by detector biases. It is thus reasonable to consider this process as a quasi two-body dipion decay. Hence, from Equation~\ref{eq:master} the mass shift due to the momentum scale and energy loss takes the form
\begin{equation}
  \Delta m_{\psi(2S)} = \left(\alpha + \frac{2\delta}{p_{\pi^+\pi^-}} \right) \cdot \meff_{\psi(2S)}.
   \label{eq:psi}
 \end{equation}
 Simulation studies with \textsc{RapidSim} confirm this is indeed the correct functional form and give $\meff_{\psi(2S)} = 411 \, \textrm{MeV}/c^2$.\footnote{Again, this differs significantly from the $Q$-value which is $310 \, \textrm{MeV}/c^2$.} As an example,  in Figure~\ref{fig6}(a), simulated $\psi(2S) \rightarrow J/\psi \pi^+ \pi^-$ decays with $\alpha = 2 \times 10^{-4}$ and $\delta = 2 \, \textrm{MeV}/c$ are compared to the expectation from Equation~\ref{eq:psi}. It can be seen that agreement is excellent. A similar procedure could be applied to the $b-$hadron decay modes such as $B^0_s \rightarrow J/\psi \phi$ and $B^0 \rightarrow J/\psi K^*$ to also allow their use for calibration.

Though understanding biases on the opening angle is more complicated for multibody decays studies of the mass shift versus the parent momentum may still to elucidate possible biases. As an example, \textsc{RapidSim}  is used to study the effect of bias on the opening angle between the pions for  the $\psi(2S) \rightarrow J/\psi \pi^+ \pi^-$ decay. In the simulation a bias of $\Delta \theta$ is introduced such that the direction vector of the pion pair is unchanged and the direction of each pion is changed by $\Delta\theta/2$ either towards or away from that direction.  In this study an approximately linear correlation of the mass shift as a function of the dipion momentum is found (see Figure~\ref{fig6}(b) for an example). This shift can be parameterized as 
\begin{equation}
  \Delta m_{\psi(2S)} =  \frac{\meff_{\psi(2S)} \cdot p_{\pi^+\pi^-}}{4 \cdot \pstar_{eff}} \Delta \theta,
   \label{eq:psi2}
\end{equation}
where $\pstar_{eff} = 345\,$MeV$/c^2$. This indicates that as in the two-body case an observed mass shift that increases with momentum is likely to be related to a bias on the opening angle determination.
  
\section{Conclusions}
In this paper, a general formalism to quantify and correct biases on the measured invariant mass of two-body decays has been developed that is applicable at both the LHC and future facilities. It has been emphasized that insights into detector performance can be achieved by relating observed biases to physical detector parameters. In particular, incorporating understanding from these studies has the potential to improve the calibration of the momentum scale of the LHCb detector. For example, the procedure discussed in \cite{LHCb:2023ood} does not attempt to quantify biases from energy loss or vertexing by fitting the measured bias on the $J/\psi$ or $\Upsilon$ mass versus momentum. This could improve the accuracy of the calibration and hence reduce the dominant source of uncertainty for many mass measurements \cite{LHCb:2023ood}. The study here has focused on the simplest case of a two-body decay. For the case of multibody decays, linking observed mass shifts to calibration parameters is more complicated but can be done numerically with the aid of fast simulation packages such as \textsc{RapidSim}.

To illustrate the method, the example of measuring the $\Lz$ mass at LHCb has been discussed. Even with a modest sample size of $4 \times 10^7$ candidates, a statistical precision of  $0.6~\textrm{keV}/c^2$ on $m_\Lz$ can be achieved. The studies presented here indicate that systematic effects from the tracker can be controlled to $0.7~\textrm{keV}/c^2$. Consequently,  the achievable uncertainty on $m_\Lz$  will be limited to $\sim 2\,\textrm{keV}/c^2$ by the knowledge of $\MK$. This uncertainty is an order of magnitude smaller than would be achieved with the current LHCb procedure \cite{LHCb-DP-2008-001}. Achieving a $\sim 2\,\textrm{keV}/c^2$  uncertainty would represent a considerable improvement on current knowledge \cite{PDG2024} and constitute a cross-check of the E766 result with modern techniques.  In addition, comparison of the measured $\Lz$ and $\overline{\Lz}$ mass will give a $CPT$ test with at least an order of magnitude improvement in precision. Although there are certainly experimental challenges beyond those studied here, particularly related to the inhomogeneous distribution of material with respect to the decay position, the feasibility of carrying out such a study at the LHC has been demonstrated. The proposed FCC-ee collider aims to collect $3 \times10^{12}$ hadronic $Z^0$ decays \cite{Altmann:2025feg}. With a $\Lambda$ multiplicity of $0.388 \pm 0.009$ per event~\cite{PDG2024} at the $Z^0$ peak this will give a huge sample of $\Lambda$ decays for mass measurements and $CPT$ tests. The methods discussed here can be applied for calibration and to determine systematic uncertainties. The authors look forward to this study as well as the measurement of other hyperon masses being carried out in the future both by LHCb and at proposed $e^+e^-$ colliders.

\bmhead{Acknowledgements}

The authors thank W.~Barter, A.~Bohare, P.~ Kodassery and M.~Williams for useful discussions and proofreading of the manuscript. We thank F.~Blanc and V.~Vagnoni for further suggestions to improve the text. 

\noindent{}For the purpose of open access, the author has applied a Creative Commons Attribution (CC BY) licence to any Author Accepted Manuscript version arising from this submission.

\bmhead{Funding Statement}
The work of MN is supported by the Science and Technology Facilities Council (STFC) via a consolidated grant (ST/W000482/1). AC thanks the Edinburgh School of Physics and Astronomy for the award of a summer studentship that allowed her to work on this project.

\bmhead{Availability of data and materials}
The study of the sensitivity of the LHCb detector to  measure the $\Lz$ mass uses simulated data generated with \textsc{RapidSim}. Those samples and the code used to analyse them will be made available by the authors on request.

\bmhead{Author Contributions}
AC: Coding and simulation studies, YL: Coding and simulation studies, MN: Conceptualization, methodology, writing. All authors reviewed the manuscript

\bmhead{Ethics approval and consent to participate}
Not applicable.

\bmhead{Consent for publication}
All authors consent to publication

\bmhead{Competing interests}
The authors declare that they have no competing interests.


\end{document}